\documentclass[journal]{IEEEtran}
\usepackage{amsmath,amsfonts}
\usepackage{graphicx}
\usepackage{enumitem}

\usepackage{algorithm}
\usepackage{bm}
\usepackage{array}
\usepackage{mathtools}
\usepackage{color,soul}
\usepackage{siunitx}
\usepackage[table,xcdraw]{xcolor}

\def\BibTeX{{\rm B\kern-.05em{\sc i\kern-.025em b}\kern-.08em
T\kern-.1667em\lower.7ex\hbox{E}\kern-.125emX}}
\usepackage{balance} 
\usepackage{multirow}
\usepackage{cite}

\usepackage[utf8]{inputenc}

\usepackage{pgfplots}
\usepackage{breqn}
\usepackage[inkscapeformat=png]{svg}
\usepackage{stackengine}

\pgfplotsset{compat=1.14}

\DeclareMathOperator{\E}{\mathbb{E}}

\DeclareMathOperator{\En}{\mathcal{E}}

\newcommand{\T}{\text{T}} 
\newcommand{\Hm}{\text{H}}

\DeclareMathOperator{\bvx}{\textbf{x}}
\DeclareMathOperator{\bvg}{\textbf{g}}

\DeclareMathOperator{\bvy}{\textbf{y}}

\DeclareMathOperator{\bvw}{\textbf{w}}
\DeclareMathOperator{\bvz}{\textbf{z}}

\DeclareMathOperator{\s}{\textbf{s}}

\DeclareMathOperator{\Cm}{\textbf{C}}

\DeclareMathOperator{\Hyp}{\mathcal{H}}
\DeclareMathOperator{\Hmat}{\textbf{H}}
\DeclareMathOperator{\Gmat}{\textbf{G}}

\newcommand{\bt}{\bm{\theta}}

\newcommand{\fim}{\mbox{\boldmath$\cal I$}}

\makeatletter
\newcommand{\@giventhatstar}[2]{\left[#1\;\middle|\;#2\right]}
\newcommand{\@giventhatnostar}[3][]{#1[#2\;#1|\;#3#1]}
\newcommand{\giventhat}{\@ifstar\@giventhatstar\@giventhatnostar}
\newcommand{\Fourier}{\mbox{\boldmath$\cal F$}}

\makeatother

\begin{document}

\title{Packet Detection in a Filter Bank-Based Ultra-Wideband Communication System}
\author{Brian Nelson$^{*,\dagger}$, {\em Member, IEEE} and Behrouz Farhang-Boroujeny$^{*}$, {\em Life Senior Member, IEEE}
\thanks{$^*$Electrical and Computer Engineering Department, University of Utah, USA,
$^{\dagger}$Wireless Communications Research Department, Idaho National Laboratory, Idaho Falls, USA}}

\maketitle

\begin{abstract}
  Recently, filter bank multi-carrier spread spectrum (FBMC-SS) technology has been proposed for use in ultra-wideband (UWB) communication systems. It has been noted that, due to the spectral partitioning properties of the filter banks, a UWB signal can be synthesized and processed using a parallel set of signal processors operating at a moderate rate. This transceiver architecture can be used to generate UWB signals, without requiring a high-rate analog-to-digital and/or digital-to-analog converter. In this paper, beginning with a design operating on a single signal processor, we explore the development of a packet detector using the Rao score test. Taking advantage of the FBMC-SS signal structure, an effective detector design based on a cascade channelizer is proposed. We refer to this design as singe-radio band (SRB) detector. Given the typical bandwidth of UWB systems ($\bf 500$~MHz or wider), the SRB detector has to operate at a fast sampling rate of greater than $\bf 500$~MHz. This may be undesirable, as low cost analog-to-digital (ADC) and digital-to-analog (DAC) converters are often limited to a sampling rate of $\bf 200$~MHz or lower. Taking note of this point, the proposed SRB detector is  extended to a multi-radio band (MRB) detector, where a set of parallel signal processors operating at a moderate sampling rate are used for a more practical implementation of the detector.  Through computer simulations, we show that SRB and MRB detectors have the same performance in typical UWB channels. Finally, we provide results from an over-the-air demonstration of a UWB design occupying $\bf 1.28$~GHz of bandwidth. We find that reliable detection performance is possible in the harshest environments, at signal-to-noise ratios as low as $\bf -40$~dB with a preamble length of approximately half the duration of longest preamble length recommended in the IEEE802.15.4 standard.
\end{abstract}

This work has been submitted to the IEEE for possible
publication. Copyright may be transferred without notice, after
which this version may no longer be accessible.

\section{Introduction}
\IEEEPARstart{U}{ltra}-wideband (UWB) systems provide unique opportunities for integrated sensing and communications (ISAC). UWB systems use a bandwidth that is either $\ge 500$~MHz or $\ge 20$\% of the center (carrier) frequency. Due to the large signal bandwidth, UWB designs can facilitate precise sensing measurements and are currently standardized for low-rate communications and ranging applications \cite{9144691}. UWB systems have been approved for unlicensed transmission in many regions of the world, provided that the transmitted signal's power spectral density (PSD) at the antenna output remains below a specified regulatory spectral mask. In the United States, transmission is allowed between 3.1 and 10.6~GHz as long as the PSD of the transmitted signal remains below $-41.3$~dBm/MHz.

In an ISAC design, the transmitted signal serves a dual purpose: it carries information while simultaneously enabling sensing of the surrounding environment. This concept has attracted significant attention in current and next-generation wireless communication systems, particularly as location-aware applications become increasingly prevalent and desirable. To achieve precise sensing, signals with large bandwidths, such as UWB signals, are required. Moreover, since UWB systems typically operate at low carrier frequencies, they offer better propagation characteristics and channel stationarity compared to alternative ISAC designs that rely on higher carrier frequencies to obtain sufficient bandwidth \cite{ChengXiang2024IMSI,ChengXiang2022CNaC}.  While massive MIMO (mMIMO) architectures provide desirable values of diversity and processing gain, they often entail increased transceiver complexity and cost \cite{LiuFan2022ISaC}. Therefore, UWB-based designs present unique opportunities for ISAC systems by balancing performance, complexity, and cost.   

Although UWB offers unique opportunities for ISAC designs, UWB devices operate within the same spectrum as primary (licensed) users. These primary users are not constrained by the UWB spectral mask and may therefore transmit at significantly higher PSD levels, even though each occupies only a narrow spectral band. As a result, UWB signals are highly susceptible to partial-band interference \cite{nelson2024_int_surv}.
In addition to this interference, it has been observed in \cite{NelsonBrian2024FfUf} that the received UWB signal power can fall well below the receiver noise floor. This effect is particularly severe in non-line-of-sight (NLOS) scenarios, where the signal-to-noise ratio (SNR) may degrade to as low as $-30$ to $-40$~dB. Furthermore, the  ultra-wide bandwidth enables the receiver to resolve many multipath components of the channel, necessitating advanced equalization and combining techniques to coherently collect the received energy \cite{WinM.Z.1998Otec,NelsonBrian2024FfUf}.
These challenges are further compounded by the sampling bottleneck in legacy UWB systems, where the extremely high sampling rates required for digital processing place stringent demands on receiver hardware and analog-to-digital converters (ADCs).

Recently, filter bank multi-carrier spread spectrum (FBMC-SS) designs have been proposed for application in UWB systems \cite{UWB2024}. In a FBMC-SS framework, data symbols are spread across multiple narrow subcarrier bands, providing both frequency diversity and processing gain \cite{DarylUCC}. Owing to its robustness against partial-band interference and low-complexity channel equalization, FBMC-SS represents a promising candidate for addressing key challenges inherent to UWB-based ISAC systems. Furthermore, \cite{UWB2024} introduced a transceiver architecture that employs a bank of parallel, low-rate FBMC-SS transceivers, each tuned to an adjacent portion of the spectrum, thereby realizing a composite UWB transceiver. This architecture allows the development of a UWB system with an arbitrarily large effective bandwidth, constructed from a collection of power-efficient, low-rate transceiver modules. In this work, we consider two implementation cases of UWB (i)
a single-radio-band (SRB) configuration, where a single radio chain is employed for signal processing, and (ii)~a multi-radio-band (MRB) configuration, in which multiple radio chains are tuned to different portions of the spectrum and jointly process the transmitted and received signals.

In prior studies, \cite{UWB2024,NelsonBrian2024FfUf}, FBMC-SS UWB architectures have been investigated primarily in terms of system design and performance, without explicit consideration of synchronization. However, in UWB systems, packet detection and synchronization present significant challenges due to the severe multipath propagation, intense interference environment, and extremely low signal-to-noise ratios (SNRs). In this paper, we address this gap by developing a packet detection algorithm tailored for FBMC-SS based UWB systems, designed to operate robustly under the harsh propagation and interference conditions characteristic of UWB environments.

In \cite{Taylor_NMF}, a novel detection framework termed the {\em normalized matched filter} (NMF) was proposed, leveraging the near-perfect reconstruction fast convolution filter banks (FC-FB) to detect FBMC-SS signals under harsh interference conditions. In this design, for each FC-FB processing block, a coarse estimate of the received signal power within each subcarrier band is used to normalize the signal power across each subcarrier band, thereby suppressing narrowband interferers in a blind manner. Although this approach demonstrated exceptional detection performance, it was developed under the assumption of a single-path channel. While this assumption may hold approximately for scenarios with short delay spreads, it becomes less accurate in environments characterized by extended multipath propagation, such as those encountered in UWB systems.
In \cite{KayStevenM.1998Foss}, a detector was derived for chirp signals operating in a non-white, unknown interference environment. Although that approach achieved high detection performance for wideband signals, it does not account for multipath channel effects nor does it support a low-rate, parallelizable implementation, both of which are essential for practical UWB receiver designs.

In this paper, we develop a new solution for the detection problem by utilizing the framework of the Rao score test, \cite{KayStevenM.1998Foss}. By exploiting the structured nature of FBMC-SS signals, we derive a simplified form of the Rao score test tailored for this signal class. By leveraging both the FBMC-SS signal structure and the simplified test statistic, we propose a hardware-efficient packet detector architecture based on a cascade filter bank channelizer \cite{HarrisFredFredricJ.2021Mspf}.
The proposed design is applicable to both SRB and MRB configurations. Importantly, for channels exhibiting long multipath delay spread, as is commonly encountered in the UWB environment, the proposed detector maintains comparable performance across both SRB and MRB implementations.

Excellent performance of the proposed detector is demonstrated through computer simulations as well as by capturing the signals transmitted over-the-air through a UWB hardware prototype. For simulations, we use the standardized channel model proposed in the IEEE802.15.4 UWB standard documents. The over-the-air signals are generated using an eight channel MRB hardware. Each channel has a bandwidth of $160$~MHz, leading to a synthesized signal with a total bandwidth of $1.28$~GHz.

The remainder of this paper is organized as follows. In Section~\ref{sec:system_model}, we present our system model and a brief overview of the overlapped filtered multi-tone spread spectrum (OFMT-SS) waveform, as an example of the broad class of FBMC-SS waveforms. In Section~\ref{sec:detector_design}, we present a review of the relevant concepts from the detection theory and composite hypothesis testing and make use of them to derive the Rao score test for the problem of interest in this paper. In Section~\ref{sec:architecture}, we develop the system architecture for calculating this Rao score test in the context of a SRB detector.  In Section~\ref{sec:multi_channel}, we extend the design to a MRB detector. In Section~\ref{sec:performance}, we compare the performance of the SRB and MRB detectors. In Section~\ref{sec:cfo}, we explore the impacts of carrier frequency offset (CFO) on the detector. In Section~\ref{sec:simulations}, computer simulations are used to examine the performance of the proposed detectors and benchmark them against the theoretical results.  In Section~\ref{sec:over_the_air}, our results from an over-the-air experiment are presented. In Section~\ref{sec:conclusion}, we conclude. 

\section{System Model}\label{sec:system_model}
As a specific example of a FBMC-SS waveform, we focus on the detection of an overlapped filtered multi-tone spread spectrum (OFMT-SS) packet based on a known preamble. This choice is motivated by two primary reasons. First, OFMT-SS is particularly well-suited to the UWB communication applications emphasized in this work \cite{NelsonBrian2024FfUf}. Second, OFMT-SS provides a relatively simple construction, allowing for a clear and tractable analytical formulation \cite{ofmt_paper}. Nevertheless, the detection techniques developed in this paper are not limited to OFMT-SS and can be readily extended to other FBMC-SS waveform types, such as filtered multi-tone spread spectrum (FMT-SS) modulation \cite{DarylUCC} and staggered multi-tone spread spectrum (SMT-SS) modulation \cite{austin_smt_channel_estimation}.

In OFMT-SS, a preamble signal at baseband, i.e., prior to modulation to a carrier for transmission, is given as
\begin{equation}
  s(t) = \sum_{n=0}^{N-1}  s[n] g(t- nT_b),
\end{equation}
where $N$ is the number of symbols in the preamble, $L$ is the number of subcarrier bands, $s[n]$ is the preamble symbol sequence, and $T_b$ is the symbol duration. In addition,
\begin{equation}
  g(t) = \sum_{k=0}^{L-1}\gamma_k h_k(t),
\end{equation}
where $\gamma_k$ is the spreading gain for the $k^{\text{th}}$ subcarrier band, $h_k(t) = h(t)e^{j2\pi f_k t}$ is the $k$th subcarrier pulse shaping filter centered at the frequency $f_k$, and $h(t)$ is a symmetric square root Nyquist filter. Also, 
\begin{equation}
  f_k=\frac{k-\frac{L + 1}{2}}{T_b}, ~~~\mbox{for }k=0, 1, \cdots, L-1.
\end{equation}
Note that, here, the subcarrier bands are spaced at integer multiples of the symbol rate $f_b=1/T_b$. Moreover, in \cite{ofmt_paper}, it is shown by choosing the spreading coefficients as $\gamma_k = j^k \zeta_k$ where $j=\sqrt{-1}$ and $\zeta_k \in \left\{\pm 1 \right\}$, the combined transmitter filter  and receiver matched filter impulse response $\rho(t) = g(t)\star g^*(-t)$ is a Nyquist pulse with a main lobe width of $2T_b/L$. This leads to a signal with a flat PSD across the band of transmission. This, clearly, is an ideal choice for making the best use of the spectral band (i.e., maximizing the transmit power) while satisfying the spectral mask.

After passing through a multi-path channel, the received signal will be given by
\begin{equation}
  x(t) = c(t) \star s(t) + w(t),
\end{equation}
where
\begin{equation}
c(t) = \sum_{i} c_i\delta(t - \tau_i)
\end{equation}
is the complex-valued baseband channel impulse response and $w(t)$ is the channel noise. We assume the channel remains stationary over the length of the preamble. We assume that $w(t)$ is a zero mean wide sense stationary (WSS) circularly symmetric complex Gaussian random process with autocorrelation function $R_{ww}(\tau)$.

At the receiver, $y(t)$ is match filtered with the conjugate, time reversed transmit filter $g^*(-t)$ to arrive at 
\begin{equation} \label{eq:mf}
  y(t) = \sum_{n=0}^{N-1}s[n]c(t) \star \rho(t - nT_b) + g^*(-t)\star w(t).
\end{equation}
Considering the signal samples at the interval  $T_s = \frac{T_b}{L}$, \eqref{eq:mf} converts to
\begin{equation} \label{eq:sampled_rx}
  y[m] = \sum_{n=0}^{N-1}s[n] \theta[m - nL] + w[m]
\end{equation}
where
\begin{equation}\label{eqn:theta[l]}
  \theta[l] = \sum_{i} c_i\rho\left(l T_s - \tau_i\right), ~~\text{for } 0 \le l \le p - 1
\end{equation}
and
\begin{equation}
w[m] = \left. g^*(-t)\star w(t) \right|_{t=mT_s}.
\end{equation}
In \eqref{eqn:theta[l]}, $p$ refers to the number of samples in the channel delay spread. In the developments that follow, it is assumed that $p<L$.

Next, we form the vectors 
\begin{align}
  \s &= \begin{bmatrix} s[0] & s[1] & \cdots & s[N - 1] \end{bmatrix}^\T, \nonumber \\
  \bvy &= \begin{bmatrix} y[0] & y[1] & \cdots & y[NL - 1] \end{bmatrix}^\T, \nonumber \\
  \bvw &= \begin{bmatrix} w[0] & w[1] & \cdots & w[NL - 1] \end{bmatrix}^\T, \nonumber \\
  {\bt} &= \begin{bmatrix} {\theta}[0] &{\theta}[1] & \cdots &{\theta}[p - 1] \end{bmatrix}^\T,
\end{align}
and the matrix
\begin{equation} \label{eq:H_mat_def}
  \Hmat = \s \otimes \textbf{P},
\end{equation}
where $\otimes$ denotes the Kronecker product, 
\begin{equation} \label{eq:p_mat}
  \textbf{P} = \begin{bmatrix}
    \textbf{I}_p \\
    \textbf{0}_{L-p, p}
  \end{bmatrix},
\end{equation}
$\textbf{I}_p$ is the identity matrix of size $p$, and $\textbf{0}_{L-p, p}$ is the $(L-p)\times p$ matrix of zeros. Using these definitions, the received signal can be formulated in the classical linear model
\begin{equation} \label{eq:lin_sys_model}
  \bvy = \Hmat \bt + \bvw,
\end{equation}
where $\bvw$ is the channel noise (plus interference) vector. We assume that  $\bvw\sim \mathcal{CN}(\textbf{0}, \Cm_{\bvw})$, where $\Cm_{\bvw}$ is the covariance matrix of $\bvw$. We also note that $\Hmat$ is constructed according to \eqref{eq:H_mat_def} and is only dependent on the preamble symbols $s[0]$, $s[1]$, $\cdots$, $s[N-1]$, which are assumed to be known to the receiver. Hence,  $\Hmat$ in the system model \eqref{eq:lin_sys_model} should be treated as a known coefficient matrix.

For the system model in \eqref{eq:lin_sys_model} and a given  $\Cm_{\bvw}$,  it is known that the Fisher information matrix (FIM) of $\bt$ is given by, \cite[pg. 529--530]{KayStevenM.1993Foss},
\begin{equation}\label{eqn:FIM}
\fim(\bt)=\Hmat^\Hm \Cm_{\bvw}^{-1} \Hmat.
\end{equation}
Given the form of $\Hmat$, in Appendix~\ref{app:FIM} it is shown that for the cases that the size $NL$ of the observation vector $\bvy$ is large, $\fim(\bt)$ can be approximated as
\begin{equation}\label{eqn:FIM2}
\fim(\bt)\approx \beta\textbf{I}_p
\end{equation}
where
\begin{equation}\label{eqn:beta}
\beta=\frac {N}{L}\sum_{k=0}^{L-1}\frac{1}{\Phi_{\bvw}[k]}.
\end{equation}
In \eqref{eqn:beta}, the terms $\Phi_{\bvw}[k]$ are samples of the power spectral density of the channel noise $w[n]$ at a regular spacing.

\section{Detector Design}\label{sec:detector_design}
Packet detection can be formulated as a binary hypothesis testing problem, where a decision is made between two hypotheses; the null hypothesis $\Hyp_0$ and the alternative hypothesis $\Hyp_1$. In the case here, we can formulate the detection problem as
\begin{equation} \label{eq:hyp_test}
\begin{array}{cc}  \Hyp_0: &  \bt = \textbf{0}  \\
  \Hyp_1: &  \bt \ne \textbf{0}. \end{array} 
\end{equation}

\subsection{Neyman-Pearson Detector}
When the probability density functions (PDFs) of the underlying processes are known, the Neyman-Pearson detector can be directly applied to implement an optimal detector \cite[pg. 65]{KayStevenM.1998Foss}. This detector decides $\Hyp_1$ if
\begin{equation} \label{eq:neyman_pearson}
  L(\bvy) =\frac{p(\bvy; \Hyp_1)}{p(\bvy; \Hyp_0)} > \gamma',
\end{equation} 
where $L(\bvy)$ is the likelihood function and the threshold $\gamma'$ is found from
\begin{equation}
  \int_{\left\{\bvy:L(\bvy)>\gamma'\right\}}p(\bvy; \Hyp_0) d\bvy = P_{\rm FA}
\end{equation}
where $P_{\rm FA}$ denotes the probability of false alarm.

Direct application of the Neyman-Pearson detector results in the generalized matched filter, where the noise is whitened prior to applying the matched filter while accounting for the noise whitening operation. For the signal model in \eqref{eq:lin_sys_model}, the detection performance, $P_D$, for a fixed $P_{\rm FA}$ is obtained as
\begin{equation} \label{eq:mf_bound}
  P_D = Q(Q^{-1}(P_{\rm FA}) - \sqrt{d^2}),
\end{equation}
where 
\begin{align}\label{eqn:d2}
d^2&=2 \bt^{\Hm}\Hmat^\Hm \Cm_{\bvw}^{-1}\Hmat\bt\nonumber\\
&=2\bt^{\Hm}\fim(\bt)\bt 
\end{align}
is called the {\em deflection coefficient} \cite[pg. 495]{KayStevenM.1998Foss}, and $Q(x)$ is the Gaussian right tail probability function. In addition, substituting \eqref{eqn:FIM2} in \eqref{eqn:d2}, we get
\begin{equation}\label{eqn:d2approx}
d^2\approx 2\beta\bt^\Hm \bt.
\end{equation}
Furthermore, one may take note that in \eqref{eqn:d2approx},  $\bt^\Hm \bt$ is the total power of the channel impulse response across its multi-paths. 

\subsection{Composite Hypothesis Testing} \label{ssec:composite_hyp}
In practice, the Neyman Pearson detector and its performance may not be realizable, due to unknown parameters in the PDFs under the null and alternative hypotheses. In the proposed signal model, there are a number of unknown parameters including the unknown channel gains, $\bt$, the unknown noise covariance matrix, $\Cm_{\bvw}$, an unknown carrier frequency offset, and an unknown signal start time. Because the unknown noise parameters, namely, the elements of $\Cm_{\bvw}$, are present under both the null and alternative hypotheses, they are often referred to as {\em nuisance parameters.} The unknown CFO and sample time also deserve special attention. The unknown sample time is addressed by examining the incoming signal at a sufficiently regular intervals to make sure that the packet preamble always falls within a window of the captured signal samples. The problem of CFO is addressed in Section~\ref{sec:cfo}. In the remaining parts of this section, we assume there is no CFO and  turn our attention to the theory of composite hypothesis testing \cite{KayStevenM.1998Foss}.

Composite hypothesis tests provide an analytical framework for deriving a hypothesis test when one or more parameters of the underlying processes are unknown. A number of composite hypothesis testing approaches exist for evaluating problems with the form of \eqref{eq:hyp_test}. These include generalized likelihood tests (GLRTs) and Bayesian approaches. As noted in \cite[pg. 198]{KayStevenM.1998Foss}, Bayesian approaches require multidimensional integration and more restrictive assumptions. The GLRT on the other hand, evaluates the Neyman Pearson detector at the maximum likelihood estimate (MLE) of the unknown parameters. That is, it uses the likelihood test
\begin{equation}
  L_G(\bvy) = \frac{p\left(\bvy; \hat{\bt}, \hat\Cm_{\bvw}, \Hyp_1\right)}{p\left(\bvy; \textbf{0}, \hat\Cm_{\bvw}, \Hyp_0\right)} > \gamma,
\end{equation}
where the hat signs refer to the MLE of the indicated parameters under  the respective  hypotheses,  $\Hyp_0$ or $\Hyp_1$. To select the threshold $\gamma$ and evaluate the performance of the likelihood test $L_G(\bvy) $, we make use of the asymptotic distribution of the test statistic, given by 
\begin{equation} \label{eq:glrt_asymptotic_stats}
  2 \ln L_G(\bvy) \stackrel{a}{\sim}\begin{cases}
    \chi^2_{2p}             & \text{under } \Hyp_0 \\
  {\chi'}^{2}_{2p}(\lambda) & \text{under } \Hyp_1, \\
  \end{cases}
\end{equation}
where $\chi^2_{2p}$ and ${\chi'}^{2}_{2p}(\lambda)$ are central and non-central Chi-squared distributions, respectively, each with $2p$ degrees of freedom; see \cite[pg. 205]{KayStevenM.1998Foss} for details. Here, the number of degrees of freedom is twice the number of unknown channel impulse response samples, given that these samples are complex-valued.

Using \eqref{eq:glrt_asymptotic_stats}, the threshold $\gamma$  that results in the probability of false alarm $P_{\rm FA}$ is obtained as
\begin{equation}\label{eqn:gamma}
  \gamma = Q^{-1}_{\chi^2_{2p}}(P_{\rm FA}).
\end{equation}
This is obtained by making use of $P_{\rm FA}=Q_{\chi^2_{2p}}(\gamma)$ where $Q_{\chi^2_{2p}}(\cdot)$ is the  right-tail probability of a Chi-squared random variable with $2p$ degrees of freedom. The detection probability is then
\begin{equation} \label{eq:det_prob}
  P_D = Q_{{\chi'}^{2}_{2p}(\lambda)}(\gamma).
\end{equation}

The GLRT requires evaluation of an {\em unrestricted MLE,} which is a joint MLE of the signal model parameters and the nuisance parameters. For the considered signal model, joint evaluation of the MLE under $\Hyp_1$ is difficult. An alternative test with the same performance as $LN\rightarrow\infty$ is the Rao score test. Unlike the GLRT, the Rao score test only requires evaluating the {\em restricted MLE} of the nuisance parameters under the null hypothesis, instead of evaluating the unrestricted MLE of all parameters under the alternative hypothesis. For this detection problem, this results in a considerable complexity reduction. Given that FBMC-SS designs are spread spectrum systems, $LN$ will usually be sufficiently large to approximate the performance as its asymptotic form. Due to the information orthogonality of the unknown noise covariance matrix and channel impulse response parameters, the Rao score test can be greatly simplified. Following similar procedures to \cite[pg. 375--376]{KayStevenM.1998Foss} and \cite{7575639}, it can be shown that the Rao score test can be written
\begin{equation} \label{eq:compact_rao_test_stat_text}
  T_R(\bvy) = 2 \bvy^\Hm\hat\Cm_{\bvw}^{-1}\Hmat \left(\Hmat^\Hm \hat\Cm_{\bvw}^{-1}\Hmat\right)^{-1}\Hmat^\Hm \hat\Cm_{\bvw}^{-1}\bvy.
\end{equation}
Making use of the results in \eqref{eqn:FIM}, \eqref{eqn:FIM2}, and \eqref{eqn:beta}, this result can be simplified to
\begin{equation} \label{eq:low_complexity_ts}
   T_R(\bvy) \approx \frac 2\beta \|\Hmat^\Hm \hat\Cm_{\bvw}^{-1}\bvy\|^2.
\end{equation}

\subsection{Insight}\label{ssec:insights}
To gain some insight to the Rao score test  \eqref{eq:low_complexity_ts}, we assume  $\hat\Cm_{\bvw}=\Cm_{\bvw}$, substitute \eqref{eq:lin_sys_model} in \eqref{eq:low_complexity_ts}, and make use of \eqref{eqn:FIM2}, to obtain
\begin{align} \label{eq:low_complexity_ts3}
   T_R(\bvy) \approx
    \|\sqrt{2\beta}\bt+\bvw'\|^2,
\end{align}
where $\bvw'=\sqrt{\frac 2{\beta}}\Hmat^\Hm \Cm_{\bvw}^{-1}\bvw$. We also note that $\bvw'$ is a transformed version of the noise vector $\bvw$ with the covariance matrix
\begin{align}\label{eqn:Cw'}
\Cm_{\bvw'}&=\frac 2{\beta}  \Hmat^\Hm \Cm_{\bvw}^{-1}\Cm_{\bvw}\Cm_{\bvw}^{-1}\Hmat \nonumber\\
&=2{\bf I}_p
\end{align} 
where the second line follows by recalling \eqref{eqn:FIM} and the subsequent results.

As one would expect, the result in \eqref{eq:low_complexity_ts3} is consistent with the test statistic \eqref{eq:glrt_asymptotic_stats}. Under $\Hyp_0$, where $\bt={\bf 0}$, the Rao score test $T_R(\bvy)$ reduces to $\|\bvw'\|^2$. Considering \eqref{eqn:Cw'}, this is a  random variable with $\chi_{2p}^2$ distribution. Under $\Hyp_1$, on the other hand, $T_R(\bvy)$ belongs to a ${\chi'}^2_{2p}(\lambda)$ distribution  with $\lambda=\|\sqrt{2\beta}\bt\|^2=2\beta\bt^\Hm\bt$.

It is also worth noting the impact of the length of vector $\bt$, i.e., parameter $p$, on the performance of the proposed detector. As $p$ increases, the received signal power, characterized by the vector $\bt$, is spread over a longer period of time and accordingly, in \eqref{eq:low_complexity_ts3}, mixes with a larger number of noise samples. This introduces more ambiguity in the Rao score test, hence, the detector performance degrades as $p$ increases.  To quantify this degradation, the following steps may be taken. 

To simplify our derivation here, we assume the channel noise is white and there is no interference. Under this assumption, in the system model \eqref{eq:lin_sys_model}, $\Cm_{\textbf{w}} = N_0 \textbf{I}_{LN}$, where $\frac{N_0}{2}$ is the PSD of the additive white channel noise. We also define the signal-to-noise ratio in the system model  \eqref{eq:lin_sys_model} as
\begin{equation}\label{eq:SNR}
\eta=\frac{ \bt^\Hm \Hmat^\Hm \Hmat \bt}{\E[\bvw^\Hm\bvw]}.
\end{equation}
This is the SNR after the received signal has passed through an analysis filter bank to extract the data symbol chips. Recalling the construction of $\Hmat$ and taking note that $\E[\bvw^\Hm\bvw]=\text{trace}[\Cm_{\textbf{w}}]=NLN_0$, \eqref{eq:SNR} simplifies to
\begin{equation}\label{eq:SNR1}
\eta=\frac{ \bt^\Hm \bt}{L N_0}.
\end{equation}
Next, taking note that  $\bt^\Hm \bt$ is the total received signal energy at the analysis filter bank outputs, in response to a unit power transmitted data symbol, one finds that $\En_c=\frac{ \bt^\Hm \bt}{L}$ is the averaged signal power at each chip at the output of the analysis filter bank. Making use of this in \eqref{eq:SNR1}, we get
\begin{equation}\label{eq:SNR_simp}
\eta=\frac{ \En_c}{N_0}.
\end{equation}
In the rest of this paper, our reference to SNR follows this definition. We also take note that under the present condition, where the channel noise is white and there is no interference, \eqref{eqn:beta} reduces to  
\begin{equation}
\beta=\frac{N}{N_0},
\end{equation}
hence, the non-centrality parameter  $\lambda=2\beta\bt^\Hm\bt$ simplifies to
\begin{equation}\label{eq:lambda}
\lambda=2NL\eta
\end{equation}

From \eqref{eq:low_complexity_ts3}, we see that the Rao score test is effectively an energy detector with the system model
\begin{equation}\label{eq:z system model}
  \bvz = \sqrt{2\beta} \bt + \bvw'.
\end{equation} 
In this model, the SNR  can be written as
\begin{align}\label{eq:SNRz}
  \eta_{\bvz
  } &= \frac{2\beta \bt^\Hm \bt}{\E\left[ {\bvw'}^{\Hm} \bvw' \right]} \nonumber\\
  &=\frac{NL\eta}{p},
\end{align}
where the second line follows by making use of \eqref{eq:lambda} and noting that  $\E\left[ {\bvw'}^{\Hm} \bvw' \right]=2p$.

At this point a comparison of the SNR values in \eqref{eq:SNR_simp} and \eqref{eq:SNRz} is instructive. Equation \eqref{eq:SNR_simp} is the SNR at the receiver input, after applying the analysis filter bank to remove any out of band noise/interference. Equation \eqref{eq:SNRz}, on the other hand, presents the SNR after combining the analysis filter bank output signals and matched filtering the result with the pilot sequence $s[0]$, $s[1]$, $\cdots$, $s[N-1]$. Here, the former leads to a processing gain $L$ and the latter results in a processing gain $N$. However, since the received signal is spread over $p$ samples in time, there exists a processing gain loss of $1/p$. This is in agreement with \eqref{eq:SNRz}, where we see a processing gain $NL/p$. This result is also in agreement with our earlier argument where it was noted that the Rao score test degrades as $p$ increases.

To gain further insight into the impact of channel uncertainty on the performance of the proposed Rao test score, we explore the loss in SNR performance required for a particular $P_D$ as the number of unknown channel parameters $p$ increases. Appendix~7A of  \cite{KayStevenM.1998Foss} presents a performance analysis of an energy detector with a system model similar to \eqref{eq:z system model}. The result there, translated to our notations here, leads to a probability of detection $P_D$, similar to \eqref{eq:mf_bound} with the deflection coefficient $d^2$ replaced by 
\begin{equation}\label{eq:d2_rao_def}
  d^2_{\text{Rao}} = \frac{(NL\eta)^2}{p}.
\end{equation}
Recognizing that for a fixed $P_{\text{FA}}$, the detection performance is specified by the deflection coefficient, we solve \eqref{eq:d2_rao_def} for $\eta$, to obtain
\begin{equation}
\eta=\frac{d_{\text{Rao}}\sqrt{p}}{NL},
\end{equation}
 or
 \begin{equation}\label{eq:snr_loss_for_longer_chan_ds}
10\log_{10}\eta=10\log_{10}\frac{d_{\text{Rao}}}{NL}+5\log_{10}p.
\end{equation}
This result shows that for a given $d_{\text{Rao}}$, i.e., a set performance of the detector, each doubling of $p$ translates to 1.5~dB increase in the required SNR.

These findings, within a good approximation, remain applicable to the case where part of the channel has been interfered. The presence of $\hat\Cm_{\bvw}^{-1}$ in the Rao score test only leads to the application of  lower weights to the interfered chips. As a result, the presence of narrow-band interferers results in some loss in the processing gain, but without any significant effect on other aspects of the detector.

Finally, we take note that computation of the Rao score test in \eqref{eq:low_complexity_ts} requires the estimate $\hat\Cm_{\bvw}$ as well as the estimates of $\Phi_{\bvw}[k]$ (for computation of $\beta$). In the next section, considering the filter bank-based construction of the underlying signals, we show how an estimate of the Rao score test $T_R(\bvy)$ can be obtained without a direct estimate of $\hat\Cm_{\bvw}$.

\section{An Implementation of Rao score test Based on a Filter Bank Structure} \label{sec:architecture}
To develop an efficient structure for computation of the Rao score test \eqref{eq:low_complexity_ts}, we note that the signal vector $\bvy$ should be first obtained by passing the received signal $x(t)$ through the matched filter  $g^*(-t)$; see \eqref{eq:mf}. With this in mind, we rewrite \eqref{eq:low_complexity_ts} as
\begin{equation} \label{eq:low_complexity_ts_freq}
   T_R(\bvy) \approx \frac 2\beta ||\Hmat^\Hm \hat\Cm_{\bvw}^{-1}\Gmat^{*}\bvx||^2
\end{equation}
where $\bvx$ is a vector of length $LN$ of samples of the received signal and $\Gmat^{*}$ is a circulant matrix realizing the matched filtering operation. One may take note that this does not result in an accurate computation of $\bvy$, because the wrap around coefficients  in $\Gmat^{*}$ distorts the elements of $\bvy$ near its beginning and its end. At this point, we ignore this loss of exactness in an effort to provide some explanation to the final structure that is presented in the sequel.

Next, from the discussion in Appendix~\ref{app:FIM}, we recall that as the data record length $NL$ tends to infinity, $\Cm_{\bvw}^{-1}=\Fourier^{\Hm}\bm{\Lambda}_{\bvw}^{-1}\Fourier$, where $\Fourier$ is the normalized DFT matrix of size $LN\times LN$ satisfying the identity $\Fourier^{\Hm}\Fourier={\bf I}$, and $\bm\Lambda_{\bvw}$ is a diagonal matrix containing the periodogram samples of $w[n]$. Also, we recall that the circulant matrix $\Gmat^{*}$ may be expanded as $\Gmat^{*}=\Fourier^{\Hm}\bm{\Lambda}_{\bvg^*}\Fourier$, \cite{1054924}. Substituting these expansions of $\Cm_{\bvw}^{-1}$ and $\Gmat^{*}$ in \eqref{eq:low_complexity_ts}, we obtain
\begin{equation} \label{eq:low_complexity_ts2}
   T_R(\bvy) \approx \frac 2\beta ||\Hmat^\Hm \Fourier^\Hm \bm\Lambda_{\bvw}^{-1}\bm\Lambda_{\bvg^*}\Fourier\bvx||^2.
\end{equation}
The steps taken to compute $T_R(\bvy)$ in \eqref{eq:low_complexity_ts2} are instructive to be noted here. These steps may be summarized as:
\begin{enumerate}
\item
At the first step, the input signal vector $\bvx$ is taken and through the operations $\bm\Lambda_{\bvg^*}\Fourier\bvx$ implements the matched filtering operation in the frequency domain. We use $\bvy_{\rm f}$ to denote the result of these operations.
\item
To obtain an estimate of the diagonal elements of $\bm\Lambda_{\bvw}$, the sum of magnitude squares of the elements of $\bvy_{\rm f}$, within each subcarrier band is taken as an estimate of the signal power in that band. This will be used to set the elements of $\bm\Lambda_{\bvw}$ at the respective subcarrier band and this is used to complete the operation $\bm\Lambda_{\bvw}^{-1}\bvy_{\rm f}=\bm\Lambda_{\bvw}^{-1}\bm\Lambda_{\bvg^*}\Fourier\bvx$.
\item
The result of Step 2) is converted to the time domain by applying $\Fourier^\Hm$.
\item
Multiplication of the result of Step 3) by $\Hmat^\Hm$ and the subsequent operations are carried out in the time domain. We also note that the signal power estimates in Step 2) are estimates of $\Phi_{\bvw}[k]$ which will be used to compute $\beta$.
\end{enumerate}

Although the above procedure may be thought of as a possible method of computation of the Rao score test without a prior knowledge of the covariance matrix $\Cm_{\bvw}$, this is not our choice in this paper but was merely presented to guide us towards a more effective structure that is introduced next. The reasons for not using the above procedure are
\begin{itemize} 
\item
The size the input vector $\bvx$ is usually very large (in the order of $10^5$ or larger). Hence, this requires a large FFT size that may not be a good choice in a practical application.
\item
Since the time of arrival of a packet is not known, the match filtering operation and the subsequent steps should be applied to a running window of incoming signal samples and the detection process should be repeated after reception of a relatively small set of signal samples. 
\end{itemize}

\begin{figure}[t]
  \centering
  \includegraphics[width=\columnwidth]{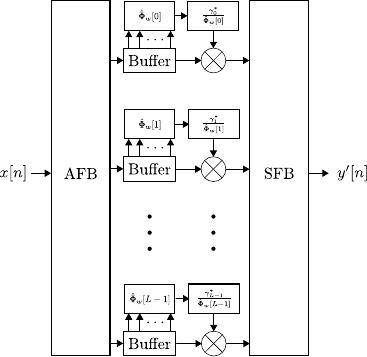}
  \caption{Cascade channelizer architecture for a SRB detector.}
  \label{fig:cascade_chan}
\end{figure}

In light of the above points and with the understanding obtained through the discussions surrounding \eqref{eq:low_complexity_ts2}, we propose the analysis/synthesis filter bank (AFB/SFB) structure in  Fig.~\ref{fig:cascade_chan} as an effective method of obtaining the elements of the vector  $\bvy'=\hat\Cm_{\bvw}^{-1}\bvy$. The estimates $\hat\Phi_w[k]$, for $k=0,1,\cdots,N-1$, are also used to calculate $\beta$ according to \eqref{eqn:beta}.  The structure in Fig.~\ref{fig:cascade_chan} is based on the non-maximally decimated perfect reconstruction cascade channelizer design of \cite{HarrisFredFredricJ.2021Mspf,XiaofeiChen2014NDAF}. In the cascade channelizer, the incoming signal $x[n]$ is passed through an analysis filter bank (AFB) with outputs generated at a spacing of $r/T_b$, where $r$ is an integer greater than one. The result is fed through a bank of first-in first-out (FIFO) buffers that keep a history of the past $rN$ samples of the outputs of the AFB. Recall that $N$ is the length of the preamble vector $\s$. The content of the FIFO buffers are used to obtain the estimates of $\Phi[k]$, signified by the hat signs. The squares of signal samples are averaged for this purpose. The scaling factor $\frac{\gamma_k^*}{\hat\Phi[k]}$ is then applied before passing the result to the SFB. Subsequently, the signal samples $y'[n]$ at the SFB output are passed through a bank of polyphase filters that implements $L$ matched filtering operations by the columns of $\Hmat$. The sum squares of the outputs from these filters, scaled by $2/\beta$, yields the Rao score test $T_R(\bvy)$.

The proposed architecture provides a few distinct advantages. First, the AFB performs prototype matched filtering while also channelizing the input signal. The channelized signal can then be used to directly estimate the unknown noise powers at each subcarrier band. This is in contrast to other schemes where auto-regressive noise parameter estimation is used \cite[pg. 354]{KayStevenM.1998Foss}, which may require many parameters to effectively estimate the narrowband signals experienced by a UWB transceiver. Alternatives to the auto-regressive parameter estimation approaches rely on windowed fast convolution processing to facilitate a noise power estimate \cite{Taylor_NMF}. This method is susceptible to some leakage of the noise PSD across the band, \cite{OFDMvsFBMC}, hence, lacks accurate identification of an interference at its respective position across the spectral band.

The second advantage of the proposed scheme lies in its straightforward design parameterization and inherent ability to accommodate varying channel delay spreads. Specifically, a longer multi-path delay spread can be addressed by simply increasing the number of matched-filter output samples included in the test statistic computation.

The third advantage of the proposed channelizer is its ability to scan and process the incoming signal in short steps. The input data buffer is refreshed every $L$ input samples, and each output update produces $L$ new samples of $y'[n]$. Recall that the vector $\bvy'$ represents a segment of length $LN$ taken from the channelizer output. Therefore, this vector is updated each time $L$ new samples of $y'[n]$  are generated. This step-by-step update mechanism enables the Rao detector to continuously scan the incoming signal and detect the presence of an incoming packet as it may arrive.

Finally, the Rao score test eliminates the need for joint estimation of the channel parameters and noise covariance matrix, simplifying the signal processing architecture. 

We note that the detector presented in Fig.~\ref{fig:cascade_chan} is that of a single-radio-band (SRB) detector. In the next section, we show how the above developments may be extended to a receiver based on a multiple-radio band (MRB) detector.

\section{Multi-Radio Band Design} \label{sec:multi_channel}
In certain applications, such as UWB communications and sensing, the overall transmission bandwidth may be too large to be practically processed by a single radio chain. To address this, the design proposed in \cite{UWB2024} partitions subsets of the FBMC-SS subcarrier bands across multiple radio processing chains. Each chain operates at a distinct carrier frequency, enabling low-rate baseband processing within its assigned sub-band. This architecture effectively eliminates the need for costly and power-inefficient high-rate ADC and DAC components. Building upon this concept, we demonstrate how the SRB design introduced in the previous section can be generalized to a MRB configuration.

We consider a system with $M$ radio bands. Let $\bvy_m$ and $\bvw_m$,  respectively, be $KN \times 1$ vectors denoting the received signal and noise vector on the $m^{\text{th}}$ radio band. We also let $K= \frac{L}{M}$ denote the number of subcarrier bands allocated to each radio band. Similar to the SRB case, we assume the channel  noise plus interference in each radio band is a proper complex Gaussian random vector with mean of zero and  covariance matrix $\Cm_{\bvw_m}$. We also use $\bt_m$ to denote the channel impulse response at the $m^{\text{th}}$ radio band and use $q=p/M$ to denote its length.

Combining the signals and the channel impulse responses from the $M$ radio bands, we get
\begin{equation} \label{eq:mc_model}
 \bar \bvy =\bar\Hmat\bar\bt + \bar\bvw, 
\end{equation}
where
\begin{align}
  \bar\bvy &= \begin{bmatrix} \bvy_0^\T & \bvy_1^\T & \cdots & \bvy_{M-1}^\T \end{bmatrix}^\T, \nonumber \\
  \bar\bvw &= \begin{bmatrix} \bvw_0^\T & \bvw_1^\T & \cdots & \bvw_{M-1}^\T \end{bmatrix}^\T, \nonumber \\
  \bar\bt &= \begin{bmatrix} \bt_{0}^\T & \bt_{1}^\T & \cdots & \bt_{ M-1}^\T \end{bmatrix}^\T,
\end{align}
and
\begin{equation} \label{eq:mb_data_mat}
  \bar\Hmat = \s\otimes {\bf Q}\otimes {\bf I}_M.
\end{equation}
Here, the matrix $\bf Q$ has the same form as $\bf P$ in \eqref{eq:p_mat} with $L$ and $p$ replaced by $K$ and $q$, respectively. We also take note that, by its construction, $\bar\Hmat$ is a block diagonal matrix with diagonal block elements $\s\otimes {\bf Q}$.

\begin{figure}[t]
  \centering
  \includegraphics[width=\columnwidth]{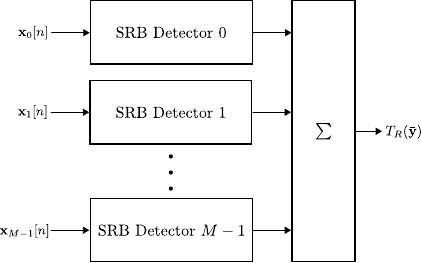}
  \caption{MRB detector architecture.}
  \label{fig:mrb_arch}
\end{figure}

To proceed, we also take note that the covariance matrix $\Cm_{\bar{\bvw}}$ of the noise vector $\bar\bvw$ is a block diagonal matrix with diagonal block elements $\Cm_{\bvw_m}$. This is because for all pairs $m\ne n$, $\bvw_m$ and $\bvw_n$ are uncorrelated with one another since they belong to different parts of the spectral band. Recalling \eqref{eqn:FIM2} and \eqref{eqn:beta} and their application  leading to \eqref{eq:low_complexity_ts}, as well as using the results in Appendix~\ref{app:FIM_mrb}, the Rao score test for the MRB system can be shown to be
\begin{align} \label{eq:simp_mc_ts}
  T_{R}(\bar\bvy) \approx  \sum_{m=0}^{M-1} \frac{2}{\beta_m}||\Hmat_m^\Hm \hat\Cm_{\bvw_m}^{-1}\bvy_m||^2,
\end{align}
where
\begin{equation}\label{eqn:betam}
\beta_m=\frac {N}{K} \sum_{k=0}^{K-1}\frac{1}{\Phi_{\bvw_m}[k]}.
\end{equation}
It is seen that the MRB test statistic can be approximated as the sum of the SRB test statistics from each radio band. This leads to the MRB detector architecture shown in Fig.~\ref{fig:mrb_arch}.

\section{Performance Comparison of SRB and MRB Detectors}\label{sec:performance}
In this section, we explore the performance differences of the SRB and MRB architectures. To this end, we make use of the asymptotic distributions of the Rao score test statistic from \eqref{eq:glrt_asymptotic_stats}. We also assume that the asymptotic approximations of the FIM used to arrive at the low complexity test statistics are accurate for both SRB and MRB detectors. From the discussion in Section~\ref{ssec:composite_hyp}, we observe that for a given false alarm probability, the detection probability is fully specified by the degrees of freedom and the non-centrality parameter $\lambda$ of the non-central $\chi^2$ distribution. When $p=Mq$, i.e., the considered channel delay spread is the same, the SRB and MRB test statistics have the same number of degrees of freedom. Therefore, any difference in performance between the SRB detector and the MRB detector will be due to  differences in the non-centrality parameters of the respective distributions.

Considering \eqref{eq:low_complexity_ts3},  one finds that, in the case of a SRB detector, the non-centrality parameter $\lambda$ is given by
\begin{equation}\label{eq:lambdaSRB1} 
 \lambda_{\text{SRB}}=2\beta\bt^\Hm\bt.
\end{equation}
Then, inserting \eqref{eqn:beta} in \eqref{eq:lambdaSRB1}, we get 
\begin{equation}\label{eq:lambdaSRB} 
  \lambda_{\text{SRB}} = \frac{2N}{L} \bt^\Hm\bt \sum_{k=0}^{L-1}\frac{1}{\Phi_{\bvw}[k]}.
\end{equation}
Similarly, one finds that
\begin{align}\label{eq:mrb_nc_param}
  \lambda_{\text{MRB}} &= \frac{2N}{K} \sum_{m=0}^{M-1} \bt_m^\Hm\bt_m \sum_{k=0}^{K-1}\frac{1}{\Phi_{\bvw_m}[k]}.
\end{align}
On the other hand, taking note that $\bt^\Hm \bt$ is the channel impulse energy over the full band of transmission, and $\bt_m^\Hm\bt_m$, for $m=0,1,\cdots, M-1$ are channel impulse response energy over a set of exclusive bands whose combinations cover the full band of transmission, we have
\begin{equation}\label{eqn:ImpulseResponseEnergy}
 \sum_{m=0}^{M-1} \bt_m^\Hm\bt_m=\bt^\Hm \bt.
\end{equation}

Next, to compare $ \lambda_{\text{SRB}}$ and $ \lambda_{\text{MRB}}$, we consider two particular scenarios. First, when the channel noise is white, $\Phi_\bvw[k]=\Phi_{\bvw_m}[k], \forall m,k$. Under this condition, making use of \eqref{eqn:ImpulseResponseEnergy} reveals that  $\lambda_{\text{SRB}}=\lambda_{\text{MRB}}$. Second, in a case where the channel consists of a large number of taps, i.e., when $q\gg 1$, one may argue variation of $\bt_m^\Hm\bt_m$ with $m$ is small and, hence, $\bt_m^\Hm\bt_m=\bt^\Hm\bt/M, \forall m$. Substituting this in \eqref{eq:mrb_nc_param}, we get
\begin{align}\label{eq:mrb_nc_param2}
  \lambda_{\text{MRB}} &= \frac{2N}{K} \sum_{m=0}^{M-1} \frac{\bt^\Hm\bt}M \sum_{k=0}^{K-1}\frac{1}{\Phi_{\bvw_m}[k]}\nonumber\\
  &=\frac{2N}{KM}\bt^\Hm\bt \sum_{m=0}^{M-1}  \sum_{k=0}^{K-1}\frac{1}{\Phi_{\bvw_m}[k]}\nonumber\\
  &=\frac{2N}{L}\bt^\Hm\bt  \sum_{k=0}^{L-1}\frac{1}{\Phi_{\bvw}[k]}= \lambda_{\text{SRB}}.
\end{align}

Beside these two particular scenarios, one may find, through a variety of numerical examples, that, in most of the cases, the difference between $ \lambda_{\text{SRB}}$ and $ \lambda_{\text{MRB}}$ remains relatively small. Hence, for most cases of practical interest there will be no loss in performance by utilizing a parallel set of lower-rate radio chains and signal processors.

These results indicate that using a non-coherent combination of radio bands in the test statistic does not incur any performance loss. This may be understood by recognizing that, due to the multi-path channel, the Rao detector must perform some non-coherent combining. The MRB design takes advantage of this fact to split the non-coherent portions of the combiner over a set of parallel, low-rate radios. When the channel delay spread is sufficiently long, hence, $p = Mq$ holds accurately, no additional uncertainty is incorporated into the detector by using a MRB design.

\section{Carrier Frequency Offset} \label{sec:cfo}
In addition to the unknown parameters specified above, a real system will also experience an unknown CFO. Unfortunately, the CFO does not fit into the Rao score test formulation because it is meaningless under the null hypothesis. In \cite{davies}, a method for handling such scenario is proposed, where the test statistic is evaluated as the supremum of a family of test statistics evaluated over the different choices of the unknown parameter(s). Practically, the test statistic is chosen as the MLE of these parameters. This may be done  in the context of the GLRT by finding the maximum test statistic over a grid of CFO candidates, \cite{KayStevenM.1998Foss}. We use $\Delta F=\{\Delta f_0,\Delta  f_1, \cdots,\Delta  f_{J-1}\}$ to denote the set of CFO grid points that we consider. Since a family of test statistics are evaluated, the overall detection performance changes. Here, we follow \cite[pg. 281-283]{KayStevenM.1998Foss} in determining the detection performance in the presence of CFO.

To determine the detection performance, we first define the $P_{\rm FA}$ as the probability of a false alarm when, under $\Hyp_0$,  at least one choice of $\Delta f_i$ results in a false alarm. That is, we let
\begin{equation}\label{eq:utc_p_fa_def}
  P_{\rm FA} = \Pr\left\{\max_{i} T_{R, i}(\bvy) > \gamma ; \Hyp_0 \right\},
\end{equation}
where $T_{R, i}(\bvy)$ is the Rao score test statistic when the CFO is set equal to $\Delta f_i$. Following a similar derivation as the one was used in \cite[pg. 282]{KayStevenM.1998Foss}, we assume the examined CFOs are sufficiently apart such that the false alarms under $\Hyp_0$ are independent of one another. This as discussed in \cite[pg. 282]{KayStevenM.1998Foss} leads to the threshold 
\begin{equation}\label{eq:gammaCFO1}
  \gamma = Q^{-1}_{\chi^2_{2p}}\left(1-(1-P_{\rm FA})^{\frac{1}{J}}\right).
\end{equation}
For cases of interest where $P_{\rm FA}$ is small, $(1-P_{\rm FA})^{\frac{1}{J}}\approx 1-\frac{P_{\rm FA}}{J}$, hence, \eqref{eq:gammaCFO1} reduces to
\begin{equation}\label{eq:gammaCFO2}
  \gamma = Q^{-1}_{\chi^2_{2p}}\left(\frac{P_{\rm FA}}{J}\right).
\end{equation}

Comparing  \eqref{eq:gammaCFO2} with \eqref{eqn:gamma}, one will find that, for a fixed $P_{\rm FA}$, the threshold $\gamma$ given by \eqref{eq:gammaCFO2} is larger than the one given by  \eqref{eqn:gamma}. Because of this change of  the threshold $\gamma$, one would expect a degradation in the detector performance. The simulation results presented in the next Section show that this loss remains small in practice.

\section{Simulations}\label{sec:simulations}
\begin{figure}[t]
  \centering
  \includegraphics[width=\columnwidth]{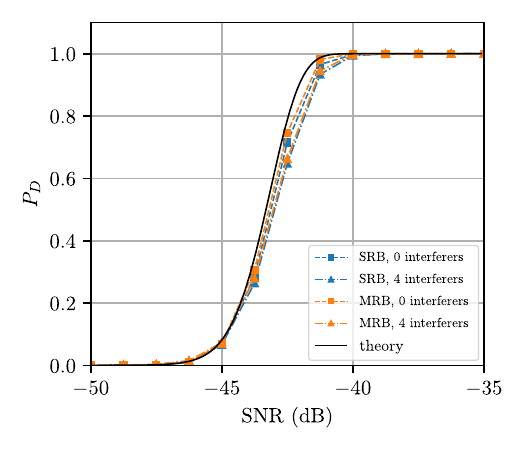}
  \caption{The detection probabilities of SRB versus MRB detectors. Theoretical curves are shown alongside the simulations for the cases where partial band interferers are present and absent. The simulated channel is the NLOS office channel.}
  \label{fig:sim_srb_v_mrb_los}
\end{figure}

\begin{figure}[t]
  \centering
  \includegraphics[width=\columnwidth]{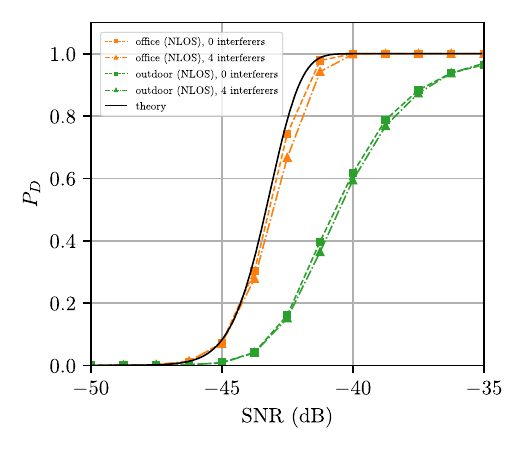}
  \caption{Simulated detection probability curves of a MRB detector in the NLOS channels of both office and outdoor environments.}
  \label{fig:sim_chan_cmp}
\end{figure}

\begin{figure}[t]
  \centering
  \includegraphics[width=\columnwidth]{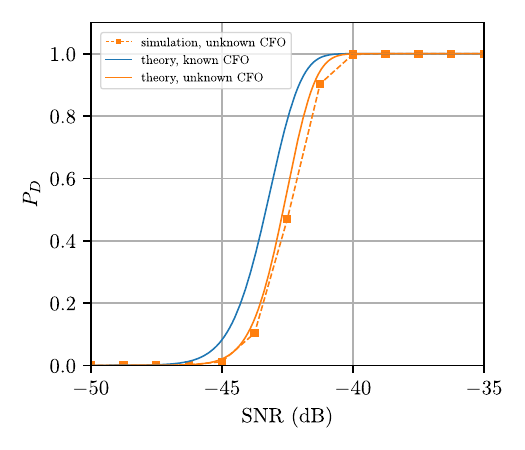}
  \caption{Simulated detection probability curves for a design with an unknown carrier frequency offset.}
  \label{fig:unknown_cfo}
\end{figure}

\begin{figure}
  \centering
  \includegraphics[width=\columnwidth]{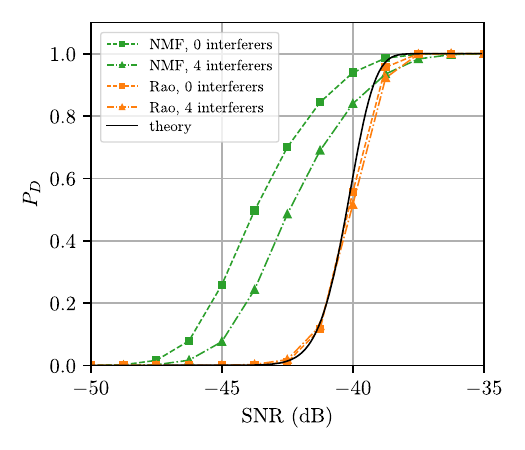}
  \caption{Simulated detection probability curves comparing the proposed Rao-based detector and the NMF for a LOS industrial channel.}
  \label{fig:nmf_cmp_los}
\end{figure}

\begin{figure}[t]
  \centering
  \includegraphics[width=\columnwidth]{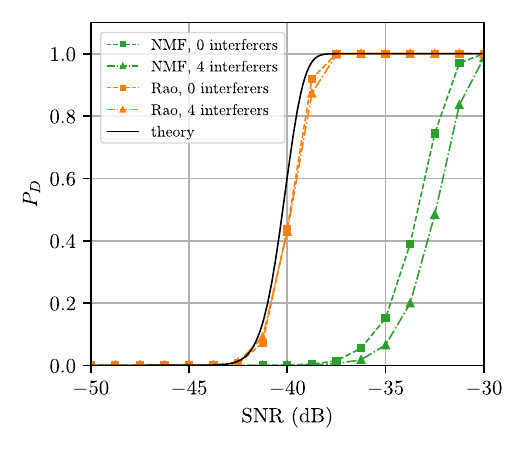}
  \caption{Simulated detection probability curves comparing the proposed Rao-based detector and the NMF for a NLOS industrial channel.}
  \label{fig:nmf_cmp_nlos}
\end{figure}

\begin{figure}[t]
  \centering
  \includegraphics[width=\columnwidth]{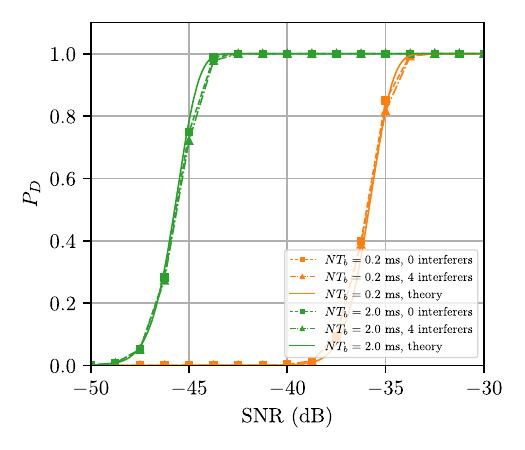}
  \caption{Detection probability curves showing the impacts of increasing bandwidth and decreasing preamble length.}
  \label{fig:hw_sim_cfg}
\end{figure}

In this section, we explore the performance of the proposed detector through a set of simulation results. Unless otherwise specified, the detectors are configured as described in Table~\ref{tab:sc_v_mc_cfg}. In this table, two base configurations are specified, a ``narrowband'' configuration operating with $500$~MHz of bandwidth and a ``wideband'' configuration operating with $1280$~MHz of bandwidth. Figs.~\ref{fig:sim_srb_v_mrb_los}--\ref{fig:nmf_cmp_nlos} are generated based on the narrowband configuration parameters. Fig.~\ref{fig:hw_sim_cfg}, on the other hand, is generated using the wideband configuration parameters. The experimental results presented in Fig.~\ref{fig:hw_curves} also make use of the wideband configuration parameters, but are limited to the choice $N=64$ only.

To provide an idea of variations of UWB channels in different environments, Table~\ref{tab:ds_params} lists typical delay spreads of UWB channels in an office, an industrial, and an outdoor setting. These are obtained by examining the outcomes of the UWB channel model simulator recommended in the IEEE802.15.4 UWB standard \cite{channel_model}. The reported durations, here, refer to the channel impulse response lengths (including the pulse shaping filter at the transmitter and matched filter at the receiver) that on average captures 95\% of the signal energy. Note that while a fixed channel impulse response length, $\tau_D$, is assumed in the implementation of the detector, the actual length of the channel may be significantly different. Some of our simulations highlight the impact of such differences.

The simulations with partial band interference use 4~interferers, each with 20~MHz of bandwidth. For each channel realization, these interferers are selected with a random power level drawn uniformly, in dB scale, from a PSD level of 5 to 40~dB above the receiver noise PSD. Additionally, the center frequencies of the interferers are drawn from a random uniform distribution covering the passband of the transmission. In \cite{nelson2024_int_surv}, it is shown that this is a realistic interference scenario for congested UWB environments. From \cite{NelsonBrian2024FfUf}, the received signal PSD in non-line of sight (NLOS) UWB links may be between $-30$ and $-40$~dB below the receiver noise level. In this SNR regime and with these interference levels, the interference PSD level is $35$ to $80$~dB above the signal PSD over the portions of the band experiencing interference. 

\begin{table}
\centering
\caption{Detector Configurations}
\label{tab:sc_v_mc_cfg}
\begin{tabular}{|c|c|c|}
\hline
\rowcolor[HTML]{C0C0C0} 
 Configuration & Narrowband & Wideband \\ \hline
 Number of subcarrier bands, $L$        & 1024 & 4096  \\ \hline
 Number of radio bands $M$              & 4    & 8     \\ \hline
 Over-the-air BW                        & $500$~MHz & $1280$~MHz \\ \hline
 Subcarrier spacing $\frac{1}{T_b}$     & $488.3$~kHz & $311.9$~kHz\\ \hline                     
 Expected channel delay spread $\tau_D$ & $80$~ns & $80$~ns \\ \hline
 False alarm probability, $P_{\text{FA}}$ & $1\times 10^{-8}$ & $1\times 10^{-8}$ \\ \hline
 Duration of preamble, $NT_b$           & $2$~ms & $2$~ms and $0.2$~ms \\ \hline
\end{tabular}
\end{table}

\begin{table}
\centering
\caption{Channel Delay Spread}
\label{tab:ds_params}
\begin{tabular}{|l|ll|}
\hline
\rowcolor[HTML]{C0C0C0} 
\cellcolor[HTML]{C0C0C0}                               & \multicolumn{2}{l|}{\cellcolor[HTML]{C0C0C0}Duration Capturing 95\% of Channel Energy (ns)} \\ \cline{2-3} 
\rowcolor[HTML]{C0C0C0} 
\multirow{-2}{*}{\cellcolor[HTML]{C0C0C0}Channel Mode} & \multicolumn{1}{c|}{\cellcolor[HTML]{C0C0C0}~~~~~~~~~~LOS~~~~~~~~~~} & \multicolumn{1}{c|}{\cellcolor[HTML]{C0C0C0}NLOS} \\ \hline
Office       & \multicolumn{1}{c|}{35} &\multicolumn{1}{c|}{47} \\ \hline
Industrial       & \multicolumn{1}{c|}{17} &\multicolumn{1}{c|}{289}  \\ \hline
Outdoor       & \multicolumn{1}{c|}{92} &\multicolumn{1}{c|}{268} \\ \hline
\end{tabular}
\end{table}

Fig.~\ref{fig:sim_srb_v_mrb_los} presents a set of sample results comparing SRB and MRB designs for the NLOS office channel. Take note that, for these simulations, the channel  delay spread $47$~ns (given in Table~\ref{tab:ds_params}) is well below the delay spread $\tau_D=80$~ns considered for the detector implementation. Under this condition, as one would expect, the simulation curves closely follow the theoretical performance, which also confirms the validity of the simplifications used to arrive at the low complexity Rao score test \eqref{eq:low_complexity_ts}.  It is also observed that the presence of interference has a minimal impact on the detector performance. Finally, we see that the MRB and SRB cases have the same performance, confirming the findings of Section~\ref{sec:performance}. With these observations and for the sake of brevity, the subsequent results are presented only for the case of the MRB detector.

Fig.~\ref{fig:sim_chan_cmp} compares the performance between the NLOS office and outdoor channels. For the outdoor NLOS channel, from Table~\ref{tab:ds_params}, we note that the delay spread $\tau_D=80$~ns used in the detector implementation is insufficient. This, as one would expect, results in some performance loss. Here, at high detection probability (close to one), the loss is $5$~dB or more.

In Fig.~\ref{fig:unknown_cfo}, the detection performance is shown for an unknown CFO over the NLOS office channel. Here, the received signal has a CFO drawn from a uniform distribution over the range  $-7$ to $+7$~kHz. The receiver computes a family of test statistics assuming different carrier frequency offsets, at 79 equally spaced positions within the latter range. The detector and its threshold are discussed in Section~\ref{sec:cfo}. Here, the CFO correlation bins are spaced such that the worst scalloping loss, \cite{HarrisF.J.1978Otuo}, will be 1~dB. As one would expect, the unknown CFO, here, introduces a detection performance loss that is less than, but close to, 1~dB.

In Fig.~\ref{fig:nmf_cmp_los} and Fig.~\ref{fig:nmf_cmp_nlos}, we compare the performance of the proposed design with the NMF in the LOS and NLOS industrial channels. In all cases, we applied a random carrier frequency offset, as was done in Fig.~\ref{fig:unknown_cfo}. For these figures, a few modifications to the configurations shown in Table~\ref{tab:sc_v_mc_cfg} are made. First, the Rao-based detector used a $\tau_D$ of 320~ns. Additionally, the NMF system was a SRB design and found its test statistic as a maximum over a window of 320~ns containing the channel impulse response.

In Fig.~\ref{fig:nmf_cmp_los}, the performance is compared in the LOS industrial channel. Here, we observe that the NMF-based detector exceeds the performance of the Rao-based detector. However, the two detectors perform about the same at the detection probabilities that are above 90\%. On the other hand,  in the case of the NLOS industrial channel, Fig.~\ref{fig:nmf_cmp_nlos}, while the Rao-based detector maintains its performance (despite the increase in channel delay spread), the NMF-based detector experiences a performance degradation of $10$~dB (or more).

In Fig.~\ref{fig:hw_sim_cfg}, we show the performance results of the proposed detector in a channel that occupies a bandwidth $1.28$~GHz. Recall that the previous results were all for a channel with a bandwidth $500$~MHz. In addition, to present some results in line with the over-the-air (OTA) results that are presented in the next section, the results in Fig.~\ref{fig:hw_sim_cfg} include the results with a much shorter preamble length (about an order of magnitude shorter). As discussed in the next section, this choice of the preamble length was forced to us by the limitation of the test environment that we used.
 
These simulation results demonstrate that high detection probabilities are possible in the harshest UWB regimes, even at a manageable sampling rate and with a moderate duration of preamble. For context, the longest preamble defined in the IEEE802.15.4 UWB standard is $4.0697$~ms \cite[pg. 649]{9144691}. Our results show that reliable detection below $-40$~dB SNR is achievable with less than half of this preamble duration.

\section{Over-the-Air Demonstration}\label{sec:over_the_air}
We have used a total of eight Ettus X310 universal software radio peripherals (USRPs) to construct a MRB UWB transceiver prototype.  Each USRP has two radio chains with a maximum sampling rate of 200 Mega samples per second and an anti-aliasing filter with passband of 160~MHz. Additionally, the sample clocks of the radios are synchronized using the Ettus OctoClock-G CDA-2990 \cite{octoclock} module. This configuration results in a total OTA bandwidth of $8\times 160 = 1280$~MHz.

\begin{figure}[t]
  \centering
  \includegraphics[width=\columnwidth]{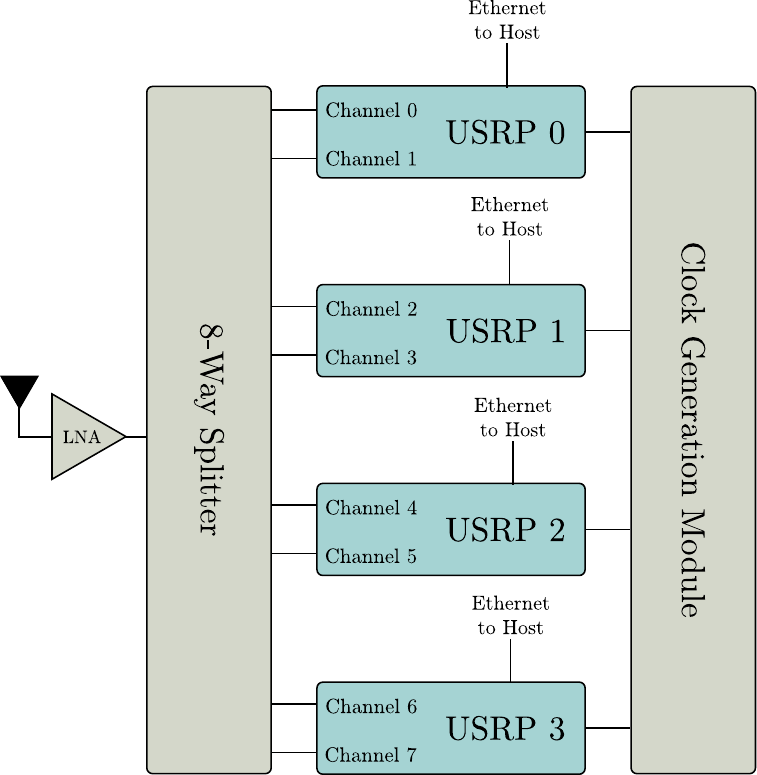}
  \caption{Diagram of the receiver hardware.}
  \label{fig:hw_diagram}
\end{figure}

We use the omni-directional Taoglas FXUWB10.01.0100C antenna~\cite{fxuwb10.01.0100C} in the transmitter and receiver. The antenna in the transmitter and receiver are at the heights of 1.21~m and 1.04~m above the floor, respectively. In the transmitter, the signal for each radio chain is synthesized on a separate radio chain and then combined using an 8-channel combiner for over-the-air transmission. This design is calibrated so that, across the full band of transmission, the signal satisfies the spectral mask specified by the federal communications commission (FCC), namely, the equivalent isotropic radiated power (EIRP) is at or below $-41.3$~dBm/MHz across the band of transmission. In the receiver, a similar architecture is used, except that a low noise amplifier (LNA) is placed before the signal splitter. This LNA compensates for the loss in signal power caused by the splitter. The radios are then connected to a transmitter and receiver host PC. Fig.~\ref{fig:hw_diagram} shows a diagram of the receiver hardware.

\begin{figure}[t]
  \centering
  \includegraphics[width=\columnwidth]{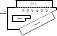}
  \caption{Diagram of the test setup showing transmitter configuration and tested locations.}
  \label{fig:floor_plan}
\end{figure}

We evaluate the performance in a NLOS office environment. To this end, we separate the transmitter and receiver by an interior wall in a small commercial office space. Fig.~\ref{fig:floor_plan} is a diagram of the test environment, where each test point is marked. To put the received signal in the context of SNR, we estimate the SNR by transmitting a preamble with a duration of 1.6~ms and calculate the power of an estimate of the multi-path channel, divided by the variance of the received signal. Table~\ref{tab:dist_to_snr} shows the estimated SNR at each distance. To explore the performance in the presence of NBI, we tuned the radios to operate from 4620~MHz to 5900~MHz, where Wi-Fi activities would act as NBI. 

\begin{table}
  \centering
  \caption{Distance to Estimated SNR}
  \label{tab:dist_to_snr}
  \begin{tabular}{|c|c|}
  \hline
  \rowcolor[HTML]{C0C0C0} 
  Distance (m) & SNR (dB)      \\ \hline
  8            & $-27.7$   \\ \hline
  10           & $-30.5$    \\ \hline
  12           & $-32.9$    \\ \hline
  14           & $-34.8$   \\ \hline
  16           & $-35.5$     \\ \hline
  18           & $-36.9$     \\ \hline
  \end{tabular}
\end{table}

\begin{figure}[t]
  \centering
  \includegraphics[width=0.92\columnwidth]{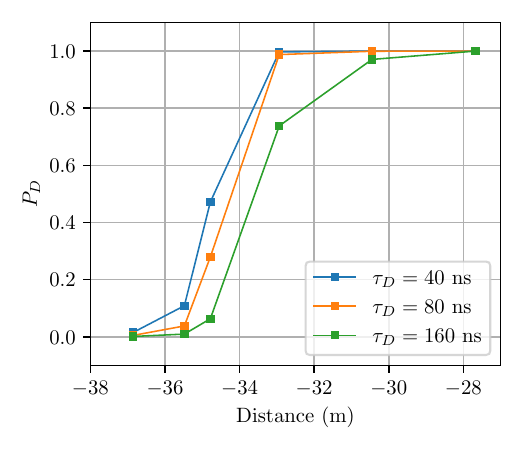}
  \caption{Over-the-air detection probability curves plotted as a function of estimated SNR for various multi-path delay window sizes.}
  \label{fig:hw_curves}
\end{figure}

Fig.~\ref{fig:hw_curves} plots the performance as a function of SNR for several choices of multi-path window size, $\tau_D$. For each case, 1,000 packets were tested. The other relevant detector parameters are summarized in the ``Wideband'' column of Table~\ref{tab:sc_v_mc_cfg}. As in the simulation section (the orange plots in Fig.~\ref{fig:hw_sim_cfg}), the preamble duration is $NT_b=0.2$~ms. As one would expect, the results, here, are close to those of the simulation results in Fig.~\ref{fig:hw_sim_cfg}; a difference of about $1$~dB is observed. This may be contributed through the fact the OTA channel may be different from the simulated channel. It is also worth noting that the choice of the preamble length, here, is to allow the transition in the detector performance curve to fall within the SNR values that our experimental system encounters, i.e., those listed in Table~\ref{tab:dist_to_snr}. 

\section{Conclusion}\label{sec:conclusion}
In this paper, we developed a packet detector for a FBMC-SS UWB system. The derivation took the harsh multi-path and interference environments likely in UWB channels into account. We found that by using the Rao score test, a low-complexity, high-performance detector could be derived. Following the theoretical derivation, a signal processing architecture based on a cascaded polyphase analysis and synthesis filter banks was presented as a method for implementing the test statistic. It was seen that the theoretical Rao detector could be extended to a MRB signal without requiring a coherent combination across radio bands. The performance of both schemes was explored and found to be approximately the same in multi-path channels of sufficient delay spreads. 

It was seen that the cascade channelizer design allowed for straightforward estimation of the nuisance parameters in the Rao score test statistic calculation, interference suppression, and prototype matched filtering. The performance of the design was shown through simulations and over-the-air demonstrations. It was shown that the simplifications and approximations used to arrive at low complexity test statistics for the SRB and MRB designs were justified due to the close match of simulation and theoretical performance. Additionally, it was demonstrated that a moderate-rate MRB design can be used without any performance loss. It was seen that this detector has excellent performance in harsh environments, even when the synchronization sequence (preamble) was a fraction of the duration of the longest preamble recommended in the IEEE802.15.4 standard.

This detector is an enabler of UWB and other wideband ISAC designs, because it facilitates a scalable architecture for performing initial receiver synchronization. It enables synchronization without requiring high sampling rates that may be required for wide enough bandwidths to facilitate precise ISAC sensing. Additionally, because it provides a robust framework for designing a packet detector across a variety of operating environments, it has applicability to other FBMC-SS systems. 

\section*{Acknowledgements}
This research made use of Idaho National Laboratory's High Performance Computing systems located at the Collaborative Computing Center and supported by the Office of Nuclear Energy of the U.S. Department of Energy and the Nuclear Science User Facilities under Contract No. DE-AC07-05ID14517.

This manuscript has in part been authored by Battelle Energy Alliance, LLC under Contract No. DE-AC07-05ID14517 with the U.S. Department of Energy. The United States Government retains and the publisher, by accepting the paper for publication, acknowledges that the United States Government retains a nonexclusive, paid-up, irrevocable, world-wide license to publish or reproduce the published form of this manuscript, or allow others to do so, for United States Government purposes. STI Number: INL/JOU-25-87738.

\begin{appendices}

\section{Simplification of FIM}\label{app:FIM}
Our derivations in this appendix make use of the following result of linear algebra. The covariance of a random process $w[n]$ when the size of the observation vector tends to infinity may be decomposed as \cite{1054924}
\begin{equation}\label{eqn:decomposition1}
\Cm_{\bvw}=\Fourier^\Hm \bm{\Lambda}\Fourier
\end{equation}
where $\Fourier$ is the normalized DFT matrix, satisfying the identity $\Fourier^\Hm\Fourier=\bf{I}$, and $\bm{\Lambda}$ is a diagonal matrix containing the periodogram  samples of $w[n]$. Alternatively, \eqref{eqn:decomposition1} may be written as
\begin{equation}\label{eqn:decomposition2}
\Cm_{\bvw}=\sum_k \lambda_k {\bf f}_k{\bf f}_k^{\rm H}
\end{equation}
where ${\bf f}_k$ are columns of $\Fourier$, and $\lambda_k$ are diagonal elements of $ \bm{\Lambda}$. Also, it is easy to show that
\begin{equation}\label{eqn:decomposition3}
\Cm_{\bvw}^{-1}=\sum_k \frac 1\lambda_k {\bf f}_k{\bf f}_k^{\rm H}
\end{equation}

Substituting \eqref{eqn:decomposition3} in \eqref{eqn:FIM}, one finds that the $l,m^{\rm th}$ element of the FIM matrix can be written as
\begin{equation}\label{eqn:FIMlm1}
  \left[ \fim(\bt)\right]_{l,m} = \frac{1}{LN} \sum_{k=0}^{NL-1}\frac{H^*_l[k]H_m[k]}{\lambda_k},
\end{equation}
where $H_l[k]$ and $H_m[k]$ are the DFTs of the $l^{\text{th}}$ and $m^{\text{th}}$ columns of $\Hmat$, respectively. As $LN\rightarrow \infty$, \eqref{eqn:FIMlm1} converts to
\begin{equation} \label{eq:asymptotic_fim_element1}
 \left[ \fim(\bt)\right]_{l,m} = \int_{0}^{1} \frac{H^*_l(f)H_m(f)}{\Phi_{\bvw}(f)} df.
\end{equation}
Here, $H_l(f)$ and $H_m(f)$ are the discrete time Fourier transforms of the $l^{\text{th}}$ and $m^{\text{th}}$ columns of $\Hmat$, respectively, and $\Phi_{\bvw}(f)$ is the power spectral density of $w[n]$.

Next, we assume that the PSD of $w[n]$ is flat across the segments of the frequency band, each of width $1/L$. With this assumption, \eqref{eq:asymptotic_fim_element1} can be rearranged as
\begin{equation}\label{eq:asymptotic_fim_element2}
 \left[ \fim(\bt)\right]_{l,m} = \sum_{k=0}^{L-1} \frac{1}{\Phi_{\bvw}[k]} \int_{\frac{k}{L}}^{\frac{k+1}{L}} H^*_l(f)H_m(f) df,
\end{equation}
where $\Phi_{\bvw}[k]$ is the sample value of $\Phi_{\bvw}(f)$ at the $k$ band.

Considering the definition of the  matrix $\Hmat$ in \eqref{eq:H_mat_def}, we see that the $k^{\text{th}}$ column of $\Hmat$ is formed by upsampling the transmitted symbol sequence by a factor of $L$ and then delaying the upsampled sequence. These imply that
\begin{equation} \label{eq:ft_col_H}
  H_l(f) = S(fL)e^{-j 2 \pi lf},
\end{equation}
and, thus,
\begin{equation} \label{eq:ft_cols_H}
  H_l^*(f)H_m(f) = |S(fL)|^2 e^{j 2 \pi (l-m) f},
\end{equation}
where $S(f)$ is the DTFT of the pilot sequence. Substituting \eqref{eq:ft_cols_H} into the integral in \eqref{eq:asymptotic_fim_element2}, and making some rearrangements of the result, we arrive at
\begin{equation}\label{eqn:FIMlm}
  \left[ \fim(\bt)\right]_{l,m} = A[l,m] 
  \int_{0}^{1}|S(f)|^2 e^{\frac{j2\pi (l-m)f}{L}}df,
\end{equation}
where
\begin{equation}\label{eqn:A1}
  A[l,m]= \frac{1}{L}\sum_{k=0}^{L-1} \frac{1}{\Phi_{\bvw}[k]}e^{j\frac{2\pi (l-m)k}{L}}.
\end{equation}
For $l=m$, \eqref{eqn:A1} reduces to
\begin{equation}\label{eqn:A2}
  A[l,l]= \frac{1}{L}\sum_{k=0}^{L-1} \frac{1}{\Phi_{\bvw}[k]}, \forall l.
\end{equation}
For choices of $l\ne m$, on the other hand, the summation in \eqref{eqn:A1} adds up a number of positive terms with different phases that are uniformly spread over the interval $0$ to $2\pi$. Statistically, this results in a set of complex values whose amplitudes are much smaller than $A_0$. Moreover, the integral term in \eqref{eqn:FIMlm} is equal to the pilot sequence energy,  i.e., $\s^\Hm\s$.  Recalling that $|s[n]|^2=1, \forall n$, we get $\s^\Hm\s=N$. The above observations, clearly,  lead to \eqref{eqn:FIM2}.

\section{Simplification of the Multi-Radio Band FIM}\label{app:FIM_mrb}
For a MRB signal, a similar approximation can be used as was described for the SRB design, which is explored in Appendix~\ref{app:FIM}. Because $\bar{\Hmat}$ and $\Cm_{\bar\bvw}$ are block diagonal, the FIM, $\bar\fim$, is block diagonal, and its inverse is block diagonal with the inverse of the blocks of $\bar\fim$ on the diagonal. Therefore, the portion of $\bar\fim^{-1}$ used in the Rao score test is given as 
\begin{equation}
  \bar\fim^{-1} = \text{diag}\left(\begin{bmatrix}
    \Hmat_0^\Hm\Cm_{\bvw_{0}}^{-1}(\theta_{s,0})\Hmat_0 \\ \vdots \\ \Hmat_{M-1}^\Hm\Cm_{\bvw_{M-1}}^{-1}(\theta_{s,M-1}) \Hmat_{M-1}
  \end{bmatrix}\right).
\end{equation}
Then, because each block matrix has the form of the previous matrix, asymptotically as the product $L'N\rightarrow\infty$, we can write
\begin{equation}
  \bar\fim^{-1} \approx \text{diag}\left(\begin{bmatrix}
    \frac{1}{\beta_0}\textbf{I}_{q} \\
    \vdots \\
    \frac{1}{\beta_{M-1}}\textbf{I}_{q}
  \end{bmatrix}\right),
\end{equation}
where
\begin{equation} \label{eq:mc_beta}
  \beta_m=\frac {N}{K} \sum_{k=0}^{K-1}\frac{1}{\Phi_{\bvw_m}[k]}.
\end{equation}
Because this matrix is diagonal, we can use a parallel architecture to implement the test statistic.

\end{appendices}

\bibliographystyle{IEEEtran}

\bibliography{IEEEabrv,ref}

\end{document}